\documentclass[preprint,showpacs,preprintnumbers,amsmath,amssymb,nofootinbib]{revtex4}
\usepackage{bbm}
\usepackage{amsfonts}
\usepackage{slashed}
\usepackage{subfigure}

\usepackage{array}

\usepackage{booktabs}
\usepackage{color, soul}
\usepackage{mathrsfs}
\usepackage{epsfig}
\usepackage{CJK}
\usepackage{graphicx}
\usepackage{dcolumn}
\usepackage{bm}
\usepackage{amsmath}
\usepackage{amssymb}

\newcommand{\p}{\partial}

\newcommand{\g}{\gamma}

\newcommand{\G}{\Gamma}

\newcommand{\si}{{\sigma}}

\let\jnfont=\rm
\def\NPB#1,{{\jnfont Nucl.\ Phys.\ B }{\bf #1},}
\def\PLB#1,{{\jnfont Phys.\ Lett.\ B }{\bf #1},}
\def\EPJC#1,{{\jnfont Eur.\ Phys.\ Jour.\ C }{\bf #1},}
\def\PRD#1,{{\jnfont Phys.\ Rev.\ D }{\bf #1},}
\def\PRL#1,{{\jnfont Phys.\ Rev.\ Lett.\ }{\bf #1},}
\def\MPLA#1,{{\jnfont Mod.\ Phys.\ Lett.\ A }{\bf #1},}
\def\JPG#1,{{\jnfont J.\ Phys.\ G}{\bf #1},}
\def\CTP#1,{{\jnfont Commun.\ Theor.\ Phys.\ }{\bf #1},}
\def\ZPC#1,{{\jnfont Z.\ Phys.\ C }{\bf #1},}
\def\JHEP#1,{{\jnfont JHEP \ }{\bf #1},}
\def\Rv{\not{\hbox{\kern-1pt $R$}}}
\def\p{\not{\hbox{\kern-3pt $p$}}}

\begin{document}

\title{The Z+photon and diphoton decays of the Higgs boson
as a joint probe of low energy SUSY models}

\author{Junjie Cao$^{1,2}$, Lei Wu$^3$, Peiwen Wu$^3$, Jin Min Yang$^3$}

\affiliation{ $^1$  Department of Physics,
        Henan Normal University, Xinxiang 453007, China \\
  $^2$ Center for High Energy Physics, Peking University,
       Beijing 100871, China \\
  $^3$ State Key Laboratory of Theoretical Physics,
      Institute of Theoretical Physics, Academia Sinica, Beijing 100190,
      China}

\begin{abstract}
In light of recent remarkable progress in Higgs search at the LHC,
we study the rare decay process $h\to Z\gamma$  and show its correlation
with the decay $h \to \gamma \gamma$ in low energy SUSY models such as CMSSM,
MSSM, NMSSM and nMSSM. Under various experimental constraints,
we scan the parameter space of each model, and present in the
allowed parameter space the SUSY predictions on the $Z \gamma$ and
$\gamma \gamma$ signal rates in the Higgs production at the LHC and future $e^+e^-$
linear colliders.  We have following observations: (i) Compared with the SM
prediction, the $Z\gamma$ and $\gamma\gamma$ signal rates in the CMSSM
are both slightly suppressed; (ii) In the MSSM, both the $Z\gamma$ and $\gamma \gamma$
rates can be either enhanced or suppressed, and in optimal case, the enhancement factors
at the LHC can reach 1.1 and 2 respectively; (iii) In the NMSSM, the $Z\gamma$ and $\gamma \gamma$ signal
rates normalized by their SM predictions are strongly correlated, and at the LHC the rates
vary from 0.2 to 2; (iv) In the nMSSM,  the $Z\gamma$ and  $\gamma \gamma$ rates are both greatly reduced.
Since the correlation behavior between the $Z\gamma $ signal and the $\gamma \gamma$ signal
is so model-dependent, it may be used to distinguish the models
in future experiments.
\end{abstract}

\pacs{14.80.Da,14.80.Ly,12.60.Jv}

\maketitle

\section{INTRODUCTION}
Based on the measurements of $\g\g$ and $ZZ^*$ channels the ATLAS and CMS collaborations
have independently provided compelling evidence for a bosonic resonance around
125-126 GeV \cite{higgs-july-ATLAS,higgs-july-CMS}. This is a great triumph for particle physics, but it also
leads to a host of new questions about the nature of the boson. So far there exist
large uncertainties in determining the rates of the two channels, and meanwhile, observation
of the boson through other signals such as $b\bar{b}$ and $\tau^+\tau^-$ channels
is still far away from becoming significant\cite{higgs-dec}. So although the preliminary
data of the LHC indicate that the boson closely resembles the Higgs boson in the
Standard Model (SM), the deficiencies of the SM itself such as gauge hierarchy problem
suggest new physics explanation of the boson. Obviously, in order to decide
the right underlying theory, the LHC should exhaust its potential to measure the decay
channels of the boson as accurately as possible in its high luminosity phase.

Among the decay modes of the Higgs-like boson $h$, the diphoton channel plays a very important
role in determining its mass, spin and parity\cite{gg-spin}. At the same time,
since the diphoton channel is mediated by loops of charged particles, it also acts as
a sensitive probe to new physics.
In fact, this feature has been widely utilized to explain the results of the ATLAS and
CMS collaborations on the inclusive diphoton signal in 2012\cite{Carena,new-scalar,new-scalar-zg,new-fermion,new-vector,new-tensor,diphoton-SUSY-1,
diphoton-SUSY-3,diphoton-THDM,diphoton-lht}, which were $1.9\pm 0.5$ and $1.56\pm 0.43$
respectively for the signal rate normalized by its SM prediction \cite{higgs-july-ATLAS,higgs-july-CMS}.
In this note, we concentrate on another decay mode $h \to Z \gamma$. In the SM, the branching ratio of
this decay is about two thirds of that for the diphoton decay, and just like the diphoton signal,
it can provide a clean final-state topology in determining the properties of the boson, such as its
mass, spin and parity\cite{zg-spin}. Moreover, since new charged particles affecting the
diphoton decay can also contribute to the $Z\gamma$ decay, the two decay modes should be correlated,
and therefore studying them in a joint way can reveal more details about the underlying physics.
Albeit the advantages, in contrast to the diphoton decay which has been intensively studied,
the $Z\gamma$ decay was paid little attention in the past.
For example, since the discovery of the boson only several works have been devoted to this decay in new physics models
such as the type-II seesaw model\cite{zg-seesaw}, the Georgi-Machacek model\cite{zg-Georgi},
the extensions of the SM by charged scalars in different $SU(2)_L$ representations\cite{zg-extended-higgs-sector}
and the SM with extra colored scalars\cite{zg-colored-scalars},
and until very recently have the CMS and ATLAS collaborations set an upper limit on the ratio
$\sigma_{Z\g}/\sigma_{Z\g}^{\rm SM}<10$ \cite{zg-exp}.
Note that although the $Z\gamma$ signal suffers from a large irreducible background
at the LHC \cite{zg-exp}, the Higgs event from the process
$e^+e^- \to Z h \to Z Z \gamma$ can be easily reconstructed
at the next generation linear collider
with  the center of mass energy around 250 GeV \cite{ILC-design-TDR},
which is very helpful in suppressing the background for such a signal.
So there is a good prospect to precisely measure
this decay in the future.

In the following, we focus on the $Z\g$ decay channel of the SM-like Higgs boson $h$ in low
energy supersymmetric models such as the Constrained Mimimal Supersymmetric Standard Model (CMSSM)\cite{CMSSM-Kane1993} ,
the Minimal Supersymmetric Standard Model (MSSM)\cite{MSSM-Fayet,MSSM-inoue}, the Next-to-Minimal
Supersymmetric Standard Model (NMSSM)\cite{higgs-mass-NMSSM-Ellwanger:2009dp} and the Nearly Minimal Supersymmetric
Standard Model (nMSSM)\cite{nmssm-l}. We investigate the $Z \gamma$ signal of the Higgs production
at the LHC and future $e^+e^-$ linear colliders, and especially, we study its
correlations with the $\gamma \gamma$ signal. As we will show below, the $Z\gamma$ signal rate
may be either enhanced or suppressed in SUSY, and its correlation
behavior is so model-dependent that it may be utilized to distinguish the
models in high luminosity phase of the LHC.

This work is organized as follows. In Section II we introduce the basic features of the
SUSY models and present some formulae relevant to our calculation. In Section III we
first discuss the effects of new charged SUSY particles on the partial decay widths
of $h\to Z\g$, then we study in a comparative way the $Z \gamma$ and $\gamma \gamma$
signal rates of the Higgs production at different colliders.
Finally, we draw the conclusions in Section IV. Various couplings used in the calculation are given in the Appendix.

\section{The models and analytic formulae}
In a low energy supersymmetric gauge theory, the explicit form of its Lagrangian is
determined by the gauge symmetry, superpotential and also soft breaking terms.
As for the four models considered in this work, their differences
mainly come from the superpotential, which is the source for the Yukawa interactions
of fermions and self interactions of scalars.

$\mathbf{MSSM~ and~ CMSSM:}$  The MSSM\cite{MSSM-Fayet,MSSM-inoue} contain two Higgs doublets
$H_u, H_d$ and it predict five physical Higgs bosons, of which two are CP-even, one is CP-odd
and two are charged. Its superpotential takes following form
\begin{eqnarray}
W^{\rm MSSM} &=& W_F+\mu \hat{H_{u}}\cdot \hat{H_{d}},
\end{eqnarray}
where $W_F$ denotes the Yukawa interaction, and its form is given by
\begin{eqnarray}
W_F &=& \overline{\hat{u}}Y_{u}\hat{Q} \cdot
\hat{H_{u}}-\overline{\hat{d}}Y_{d}\hat{Q}\cdot
\hat{H_{d}}-\overline{\hat{e}}Y_{e}\hat{L}\cdot \hat{H_{d}}.
\end{eqnarray}
After considering appropriate soft breaking terms, one can write down the Higgs potential as
\begin{eqnarray}
V^{\rm MSSM}&=&(|\mu|^2+m^2_{H_u}) |H_u^0|^2 + (|\mu|^2 + m^2_{H_d}) |H_d^0|^2\nonumber\\
&-&(B \mu  H_u^0 H_d^0 + {\rm h.c.})+ {1\over 8} (g^2_2 + g^2_1) ( |H_u^0|^2 - |H_d^0|^2 )^2 ,
\label{higgs potential}
\end{eqnarray}
where $m_{H_u}$, $m_{H_d}$ and $B$ are all soft parameters with mass dimension, terms proportional to $|\mu|^2$ come
from the $F$-term of the superpotential, and the last term comes from gauge symmetry
(so called $D$-term). This potential indicates that, after the electroweak symmetry breaking,
the $\mu$-parameter is related to the Higgs vacuum expectation value (vev) and so it
should be ${\cal{O}}(100{\rm GeV})$. But on the other hand, since $\mu$ as the only parameter with
mass dimension appears in the superpotential, its value should naturally take SUSY-preserving scale. Such a
tremendously large scale gap is usually referred as the $\mu$-problem \cite{mu-problem}.

The theoretical framework of the CMSSM\cite{CMSSM-Kane1993} is exactly same as that of the MSSM, and
the only difference between them comes from the fact that in the  general MSSM,
all soft breaking parameters are independent\cite{pmssm}, while in the CMSSM they are correlated.
Explicitly speaking, the CMSSM  assumes following universal soft breaking parameters
at SUSY breaking scale (usually chosen at the Grand Unification scale)\cite{cmssm}
\begin{equation}
\ M_{1/2} \ , \ M_0 \ , \ A_0 \ , \tan\beta \ , \ {\rm sign}(\mu),
\end{equation}
with $M_{1/2}$, $M_0$ and $A_0$ denoting gaugino mass, scalar mass and trilinear interaction coefficient respectively,
and evolves the four parameters down to weak scale to get all the soft breaking parameters of the MSSM.
In this sense, the parameter space of the CMSSM should be considered as a subset of that for the MSSM, and so is its phenomenology.

$\mathbf{NMSSM ~ and ~ nMSSM:}$ In order to solve the $\mu$-problem in the MSSM,
various singlet extensions of the MSSM were proposed in history, and among them the most well known models
include the NMSSM and the nMSSM.  The superpotentials of these two models are
respectively given by \cite{nmssm-l,nmssm-s}
\begin{eqnarray}
W^{\rm NMSSM}&=&W_{F}+\lambda\hat{S}\hat{H}_{u}\cdot\hat{H}_{d}+\frac{\kappa}{3}\hat{S}^{3},\\
W^{\rm nMSSM}&=&W_{F}+\lambda\hat{S}\hat{H}_{u}\cdot\hat{H}_{d}+\xi_{F}M^{2}_{\rm n}\hat{S}, \label{nMSSM}
\end{eqnarray}
where $\lambda$, $\kappa$ and $\xi_F$ are dimensionless parameters of order 1, and the dimensionful parameter $M_n$ may be
naturally fixed at weak scale in certain basic frameworks where the parameter is generated at a high loop level\cite{nmssm-s}.
One attractive feature of both the models comes from the fact that, after the real scalar component of $\hat{S}$ develops
a vev $\langle S \rangle$, an effective $\mu$ parameter is generated by $\mu_{\rm eff}=\lambda \langle S \rangle$,
and its value may be as low as about $100{\rm GeV}$ without conflicting with current experiments \cite{diphoton-SUSY-3}
(in contrast, the $\mu$ parameter in the MSSM must be larger than about $200 {\rm GeV}$\cite{diphoton-SUSY-3}).
Another attractive feature of the models is that the Z boson mass may be obtained with less fine tuning than the MSSM\cite{fine-tuning-NMSSM-Ellwanger:2011mu}. In the SUSY models, after the minimization of the Higgs potential,
the $Z$ boson mass is given by\cite{fine-tuning-NMSSM-Ellwanger:2011mu}
\begin{eqnarray}
\frac{M^2_{Z}}{2}=\frac{(m^2_{H_d}+\Sigma_{d})-(m^2_{H_u}+
\Sigma_{u})\tan^{2}\beta}{\tan^{2}\beta-1}-\mu^{2},
\label{minimization}
\end{eqnarray}
where $m^2_{H_d}$ and $m^2_{H_u}$ represent the soft SUSY
breaking masses of the Higgs fields, and $\Sigma_{u}$ and $\Sigma_{d}$
arise from the radiative corrections to the Higgs potential with dominant contribution
to $\Sigma_{u}$ given by
\begin{eqnarray}
\Sigma_u \sim \frac{3Y_t^2}{16\pi^2}\times m^{2}_{\tilde{t}_i}
\left( \log\frac{m^{2}_{\tilde{t}_i}}{Q^2}-1\right).
\label{rad-corr}
\end{eqnarray}
These two equations indicate that, if the individual terms on the
right hand side of Eq.~(\ref{minimization}) are comparable in
magnitude so that the observed value of $M_Z$ is obtained without
resorting to large cancelations, relatively light stops and $\mu$ of
${\cal{O}}(100{\rm GeV})$ are preferred. As far as the NMSSM and the nMSSM are concerned,
due to the Higgs self interactions, the squared mass of the SM-like Higgs boson gets an
additional contribution $\lambda^2 v^2 \sin^2 2\beta$ (compared with its MSSM expression),
and further it can be enhanced by the doublet-singlet mixing
\cite{higgs-mass-NMSSM-Miller:2003ay,higgs-mass-NMSSM-Ellwanger:2009dp}.
Consequently, predicting a $125 {\rm GeV}$ Higgs boson does not necessarily require
heavy scalar top quarks\cite{diphoton-SUSY-3}. This is very helpful in reducing the tuning.
For example, it has been shown that the fine tuning parameter $\Delta$, which is defined by
$\Delta={\rm Max}\{|\partial \ln m_Z/\partial \ln p_i^{\rm GUT}|\}$ with $p_i^{\rm GUT}$
denoting SUSY parameter at GUT scale \cite{fine-tuning-NMSSM-Ellwanger:2011mu}), may be as
low as 4 in the two models, while in the MSSM it usually exceeds 100 \cite{diphoton-SUSY-4}.

About the nMSSM, one should note that the tadpole term in Eq.(\ref{nMSSM})
only affects the Higgs masses and the minima of the scalar potential, so
the interactions in the Higgs and neutralino sectors of the nMSSM are identical to
those in the NMSSM with $\kappa =0$. This enables us to modify the package $\textsf{NMSSMTools}$ \cite{nmssm-tools}
and use it to study the phenomenology of the nMSSM \cite{nMSSM-scan-cao}. Also note that the singlino mass vanishes
at tree level and the lightest neutralino as the dark matter candidate acquires its mass through the mixing of the singlino
with Higgsinos and gauginos. In this case, the dark matter is light and singlino dominated,
and it must annihilate through exchanging a resonant light CP-odd Higgs boson to
get the correct relic density\cite{nMSSM-scan-cao}.
As a result, the SM-like Higgs boson will decay dominantly into light neutralinos or other
light Higgs bosons so that the branching fractions
of the visible decay channels like $h\to\gamma\gamma, ~b\bar{b}, ~ZZ^*(4l), ~\tau^+\tau^-$
are suppressed\cite{nMSSM-scan-cao}. This is strongly disfavored by current LHC data as shown in\cite{diphoton-SUSY-4}.
In this work, we only take the nMSSM as an example to show its peculiar behaviors
in Higgs physics (in comparison with other new SUSY models).

{\bf Formula in calculation:} In order to study the $h \to Z\g$ decay and its correlation with the
$h \to \g\g$ decay in SUSY, we define following normalized rates at the LHC and the
international linear collider (ILC)\cite{ILC-design-TDR} as
\begin{eqnarray}
R_{Z\g} & \equiv &  \frac{\sigma(pp\to h\to Z\g)}{\sigma_{\rm SM}(pp\to h\to Z\g)}
= \frac{\si_{tot}}{\si_{tot}^{\rm SM}} \frac{{\rm
Br}(h\to Z\g)}{{\rm Br}_{\rm SM}(h\to Z\g)} \simeq
\left(\frac{C_{hgg}}{C^{\rm SM}_{hgg}}\right)^2
\cdot \frac{\Gamma_{Z\g}(h)}{\Gamma^{\rm SM}_{Z\gamma}(h)}\cdot\frac{\G^{\rm SM}_{tot}(h)}{\G_{tot}(h)},\\
R_{\g\g} & \equiv &  \frac{\sigma(pp\to h\to \g\g)}{\sigma_{\rm SM}(pp\to h\to \g\g)}
= \frac{\si_{tot}}{\si_{tot}^{\rm SM}} \frac{{\rm
Br}(h\to\g\g)}{{\rm Br}_{\rm SM}(h\to\g\g)} \simeq
\left(\frac{C_{hgg}}{C^{\rm SM}_{hgg}}\right)^2
\cdot \frac{\Gamma_{\g\g}(h)}{\Gamma^{\rm SM}_{\g\g}(h)}
\cdot\frac{\G^{\rm SM}_{tot}(h)}{\G_{tot}(h)}, \\
{\cal{K}}_{Z\g} & \equiv &  \frac{\sigma(e^+e^-\to Zh\to Z Z \g)}{\sigma_{\rm SM}(e^+e^-\to Zh\to Z Z\g)}
\simeq \left( \frac{C_{hZZ}}{C^{\rm SM}_{hZZ}} \right )^2
\cdot \frac{\Gamma_{Z\g}(h)}{\Gamma^{\rm SM}_{Z\gamma}(h)}\cdot\frac{\G^{\rm SM}_{tot}(h)}{\G_{tot}(h)}, \label{ILC1} \\
{\cal{K}}_{b\bar{b}} & \equiv &  \frac{\sigma(e^+e^-\to Zh\to Z b \bar{b})}{\sigma_{\rm SM}(e^+e^-\to Zh\to Z b \bar{b})}
\simeq \left( \frac{C_{hZZ}}{C^{\rm SM}_{hZZ}} \right )^2
\cdot \frac{\Gamma_{b\bar{b}}(h)}{\Gamma^{\rm SM}_{b\bar{b}}(h)}\cdot\frac{\G^{\rm SM}_{tot}(h)}{\G_{tot}(h)}, \label{ILC2}
\end{eqnarray}
where the Higgs production  at the LHC is dominated by the gluon fusion process,
while at the ILC with $\sqrt{s}\sim 250 {\rm GeV}$, it is dominated by the $Zh$ associated production.
Here $C_{hgg}$ and $C_{hZZ}$ are the couplings of the Higgs boson to
gluons and $Z$s respectively, and $\Gamma_{Z\gamma}(h)$, $\Gamma_{\gamma\gamma}(h)$ and $\Gamma_{b\bar{b}}(h)$
are the partial widths for the decays $h\to Z\gamma$, $h\to \gamma \gamma$ and $h\to b \bar{b}$ respectively. In getting
these formulae, we neglect SUSY radiative corrections to the signals. Those corrections are expected to be few percent
given that heavy sparticles are preferred by current LHC experiments.

In SUSY, the decays $h\to Z\gamma $ and $h \to \gamma\gamma$ get new contributions from the loops mediated by
charged Higgs bosons, sfermions (including stops, sbottoms and staus)  as well as charginos. Consequently, the formula of
$\Gamma_{Z\gamma}$ and $\Gamma_{\gamma\gamma}$ are modified by
\begin{eqnarray}
\Gamma_{Z\gamma} (h) &=& \frac{G^2_F m_W^2\,\alpha\,m_h^{3}}
{64\,\pi^{4}} \left(1-\frac{m_Z^2}{m_h^2} \right)^3 \bigg| {\cal
A}_{W}^{Z\g}+ {\cal A}_{t}^{Z\g}+{\cal A}_{\tilde{f}}^{Z\g}+{\cal
A}_{H^{\pm}}^{Z\g}+{\cal
A}_{\chi^{\pm}}^{Z\g} \bigg|^2, \\
\Gamma_{\gamma \gamma} (h) &=& \frac{G_F\alpha^2
m_h^3}{128\sqrt{2}\pi^3}\bigg|{\cal A}_{W}^{\g\g}+ {\cal
A}_{t}^{\g\g}+{\cal A}_{\tilde{f}}^{\g\g}+{\cal
A}_{H^{\pm}}^{\g\g}+{\cal A}_{\chi^{\pm}}^{\g\g}\bigg|^2,
\end{eqnarray}
where  ${\cal A}_i$ ($i=W, t, \tilde{f}, H^\pm, \chi^\pm$) denote the contribution from  particle $i$
mediated loops, and their explicit expressions are listed in the Appendix.
Note that our expressions differ from those presented in \cite{Djouadi-MSSM-0503173} in two aspects.
One is that we have an overall minus sign for the new contributions ${\cal A}_{H^\pm}$, ${\cal A}_{\tilde{f}}$
and ${\cal A}_{\chi^{\pm}}$, and an additional factor 2 for the sfermion contributions. This sign difference
was also observed recently in \cite{Geng}. The other difference is that we have included in a neat way the contributions
from the loops with two particles (such as $\tilde{f}_1$ and $\tilde{f}_2$ or  $\chi_1^\pm$ and  $\chi_2^\pm$)
running in them. Such contributions were considered to be negligibly small \cite{Djouadi-MSSM-0503173}, but our results indicate that
sometimes they may play a role. Also note that in the SUSY package $\textsf{FeynHiggs}$ \cite{feynhiggs},
the decay $h \to Z \gamma$ is not calculated. In the package $\textsf{NMSSMTools}$ \cite{nmssm-tools}, this
decay is calculated only by considering
the contributions from the SM particles and the charged Higgs boson.
We improve these packages by inserting our codes for
$h \to Z \gamma$.

\section{Numerical calculation and discussions}
In our calculation, we first perform a random scan over the parameter space of
each model by considering various experimental constraints. Then for the surviving samples
we investigate the $h \to Z\gamma$ and $h \to \gamma \gamma$ decays.
Since for each unconstrained SUSY model, there are too many free parameters involved in
the calculation we make some assumptions to simplify our analysis. Our treatment of
the MSSM and the NMSSM is as follows
\begin{itemize}
\item Firstly, we note that the first two generation squarks change little the properties
of the Higgs boson, and the LHC search for SUSY particles implies that they should be
very heavy. So in our scan, we fix all soft masses and the trilinear parameters in this sector
to be 2 TeV. We checked that our conclusions are not affected by such
specific choice.
\item Secondly, since the third generation squarks affect the Higgs sector
significantly, we set free all soft parameters in this sector except that we assume
$m_{U_3}=m_{D_3}$ and $A_t = A_b$ to reduce the number of free parameters.
\item Thirdly, considering that the muon anomalous magnetic moment is sensitive to the spectrum
of scalar muons and the decay $h \to Z\gamma$ may get significant contribution from scalar
tau (stau) sector, we assume $A_{\tau}=A_{\mu}=A_e=0$, $M_{L_3}= M_{L_2}= M_{L_1}$ and
$M_{E_3}= M_{E_2}= M_{E_1}$, and treat $M_{L_3}$ and $M_{E_3}$ as free parameters. We checked
that for our considered cases, the decay $h \to Z \gamma$ is insensitive to $A_\tau$.
\item Finally, since our results are insensitive to gluino mass, we fix it at 2 TeV.
We also assume the grand unification relation $3 M_1/5 \alpha_1 = M_2/\alpha_2$ for
electroweak gaugino masses.
\end{itemize}

To sum up, for the MSSM we scan the parameters in the following regions
\begin{eqnarray}
&&1 \le \tan\beta \le 60, ~~100 {~\rm GeV} \le \mu \le 1 {~\rm TeV},
~~100 {~\rm GeV} \le M_A \le 1 {~\rm TeV}, \nonumber \\
&& 100 {~\rm GeV} \le \left(M_{Q_3},M_{U_3}\right) \le 2 {~\rm TeV},
~~ 100 {~\rm GeV} \le \left( M_{L_3},M_{E_3} \right)\le 1 {~\rm TeV}, \nonumber\\
&& -3 {~\rm TeV}\le A_t \le 3 {~\rm TeV}, ~~50 {~\rm GeV}\le M_{1} \le 500 {~\rm GeV} .
\label{MSSMscan}
\end{eqnarray}
Note that in actual calculation $\lambda$ and $\mu_{eff}$ in the NMSSM
are usually treated as independent input parameters, and for any given value of $\mu_{eff}$
the phenomenology of the NMSSM is identical to that of the MSSM with $\mu=\mu_{eff}$ in the limit
$\lambda,\kappa \to 0$ \cite{higgs-mass-NMSSM-Ellwanger:2009dp}.
This enables us to use the package NMSSMTools \cite{nmssm-tools},
which calculates various observables and also considers various experimental constraints
in the framework of the NMSSM, to study the phenomenology of the MSSM
(note that the validity of this method has been justified by the authors of the
NMSSMTools \cite{nmssm-tools}).
In our calculation we use the package NMSSMTools-3.2.4 to perform the scan for the MSSM
by setting $\lambda = \kappa = 10^{-4}$ and $A_\kappa = -10{\rm GeV}$.
Here the value of $A_{\kappa}$ is actually irrelevant to our calculation for the MSSM
as long as it is negative and satisfies $|A_\kappa| < 4 \kappa \mu/\lambda$
(in order to guarantee the squared masses of the singlet scalars to be positive)
\cite{higgs-mass-NMSSM-Ellwanger:2009dp}.

For the NMSSM, we use the package NMSSMTools-3.2.4 to scan the region in Eq.(\ref{MSSMscan}) and
also following ranges for additional parameters
\begin{eqnarray}
&&0.5 \le \lambda \le 0.7, ~~0 \le \kappa \le 0.7,
~~|A_\kappa| \le 1 {~\rm TeV}.
\end{eqnarray}
Note that in our scan, we only consider a relatively large $\lambda$. The reason is in the NMSSM,
the properties of the SM-like Higgs boson are expected to deviate significantly
from the MSSM prediction only for a sizable $\lambda$ \cite{higgs-mass-NMSSM-Ellwanger:2009dp},
and in particular, as far as $\lambda \gtrsim 0.5$ is concerned, the Higgs mass at tree level
is maximized at $\tan \beta \simeq 1$, instead of at large $\tan \beta$ in the MSSM.

As for the nMSSM, our assumptions are same as those for the MSSM except that,
in order to explain the muon anomalous magnetic moment at $2\sigma$ level,
we assume all soft SUSY breaking parameters in slepton sector to be
100 GeV \cite{nMSSM-scan-cao}.
Such a slepton mass is allowed by the LEP II bounds,
which are 99.9 GeV for the first two generations and 93.2 GeV for the third
generation \cite{LEP-slepton}. Although the LHC also gave a bound on slepton
mass by searching for the decay
$\tilde{l} \to l \tilde{\chi}_1^0  \to l + \not{\hspace{-0.15cm}E}$
\cite{Slepton-search-ATLAS,Slepton-search-CMS}, it is not applicable to the nMSSM.
The reason is that in the nMSSM, the LSP is singlino-like and the NLSP is
usually a bino-like neutralino $\tilde{\chi}_2^0$. As a result, the dominant decay
chain of a slepton is $\tilde{l} \to l \tilde{\chi}_2^0  \to l \tilde{\chi}_1^0 A \to
l \tilde{\chi}_1^0 b \bar{b} \to l + \not{\hspace{-0.15cm}E} + jets$\cite{nMSSM-scan-cao}.
Other parameters in this model are scanned in the following ranges
\begin{eqnarray}
&&0.1 \le \lambda \le 0.7, ~~|A_\lambda| \le 1 {~\rm TeV}, ~~0 \le \tilde{m}_S \le 200 {~\rm GeV},
\end{eqnarray}
where $\tilde{m}_S $ is the soft breaking mass for the singlet Higgs field. In our calculation,
we adapt the code of the NMSSMTools to the nMSSM case as done in\cite{nMSSM-scan-cao}.

For the CMSSM, we use the package NMSPEC\cite{NMSPEC} to scan following parameter space
\begin{eqnarray}
&&100 {\rm ~GeV}\leq (M_0,  M_{1/2})\leq 2 {\rm ~TeV}, ~~1\leq \tan\beta\leq 60, ~~-3 {\rm ~TeV}\leq A_0 \leq 3 {\rm ~TeV},
\end{eqnarray}
and we take the sign of $\mu$ to be positive.
Similar to the MSSM scan,
we set $\lambda = \kappa= 10^{-4}$ and $A_\kappa=-10 {\rm GeV}$ at the GUT scale
(note that the validity of the NMSPEC to study the phenomenology of the CMSSM was
emphasized by the authors of the package \cite{NMSPEC}).

Since in the CMSSM different soft breaking parameters at electroweak scale are correlated, it is expected that
its phenomenology CMSSM is only a subset of the MSSM.

In our scan we have considered various constraints on the models, which are
from vacuum stability, the LEP and LHC searches for SUSY particles and Higgs bosons,
the electroweak observables
$\epsilon_i$ and $R_b$, B physics observables such as the branching ratio
of $B \to X_s \gamma$ and the mass difference $\Delta M_s$, the dark matter
relic density and its direct detection experiments.
When imposing the constraint from a certain observable which
has experimental central value, we require SUSY to explain the observable at $2\sigma$ level.
These constraints are described in detail in \cite{Cao-2010} and have been implemented
in the package NMSSMTools-3.2.4 \cite{nmssm-tools}.
In particular, the dark matter relic density is calculated by the package
\textsf{MicrOMEGAs} \cite{micrOMEGAs}, which now acts as an important component of the NMSSMTools.

Compared with the constraints in \cite{Cao-2010}, we have following improvement in this work
\begin{itemize}
\item We require $123 {\rm GeV} \leq m_h \leq 127 {\rm GeV} $. This mass range is favored by the LHC search for Higgs boson
after considering theoretical uncertainties\cite{higgs-july-ATLAS,higgs-july-CMS}.
\item We utilize the latest result of the XENON100 experiment  to
limit the models (at $90\%$ confidence level)\cite{xenon}. We calculate the scattering cross section of the
dark matter
with nucleon with the formula presented in \cite{Cao-2010},
and we set $f_{T_{s}} = 0.020$ with $f_{T_{s}}$ denoting the strange quark content in nucleon.
\item We consider the constraints from the recent LHCb measurement of $B_s \to \mu^+ \mu^-$, which is
$ {\rm Br}(B^0_s \to \mu^+ \mu^-) = (3.2^{+1.5}_{-1.2})\times 10^{-9}$\cite{Bsmumu}. The agreement of the measurement with its SM prediction
strongly limits the combination $\tan^6 \beta/M_A^4$ in the MSSM\cite{Bsmumu-Bobeth}.
\item We also consider the constraint from the CMS search for non-SM Higgs boson from the channel $H/A \to \tau^{+} \tau^{-}$ \cite{htautau}.
This search, like $ B^0_s \to \mu^+ \mu^-$, is very power in limiting the $\tan \beta-M_A$ plane in the MSSM.
\end{itemize}

Finally, we emphasize that in our scan we require the MSSM, NMSSM and nMSSM to explain muon g-2
at $2\sigma$ level, i.e., $a_\mu^{exp} - a_\mu^{SM}= (25.5 \pm 8.0 ) \times 10^{-10}$ \cite{g-2}.
As for the CMSSM, it has long been noticed that there exists a tension to predict a 125 GeV
SM-like Higgs boson and meanwhile to explain the muon g-2
\cite{CMSSM-Higgs-muong2-Nath-1201-0520,CMSSM-Higgs-muong2-Ellis-1202-3262,CMSSM-Higgs-muong2-Arbey-1207-1348}.  In our calculation we consider the latest LHC bounds
on $M_0-M_{1/2}$ plane of CMSSM and find that under such
latest bounds the CMSSM cannot explain the muon g-2 at $2\sigma$ level.
So we do not require the CMSSM to explain the muon g-2 at $2\sigma$ level
in our analysis.

\begin{figure}[t]
\includegraphics[width=17cm]{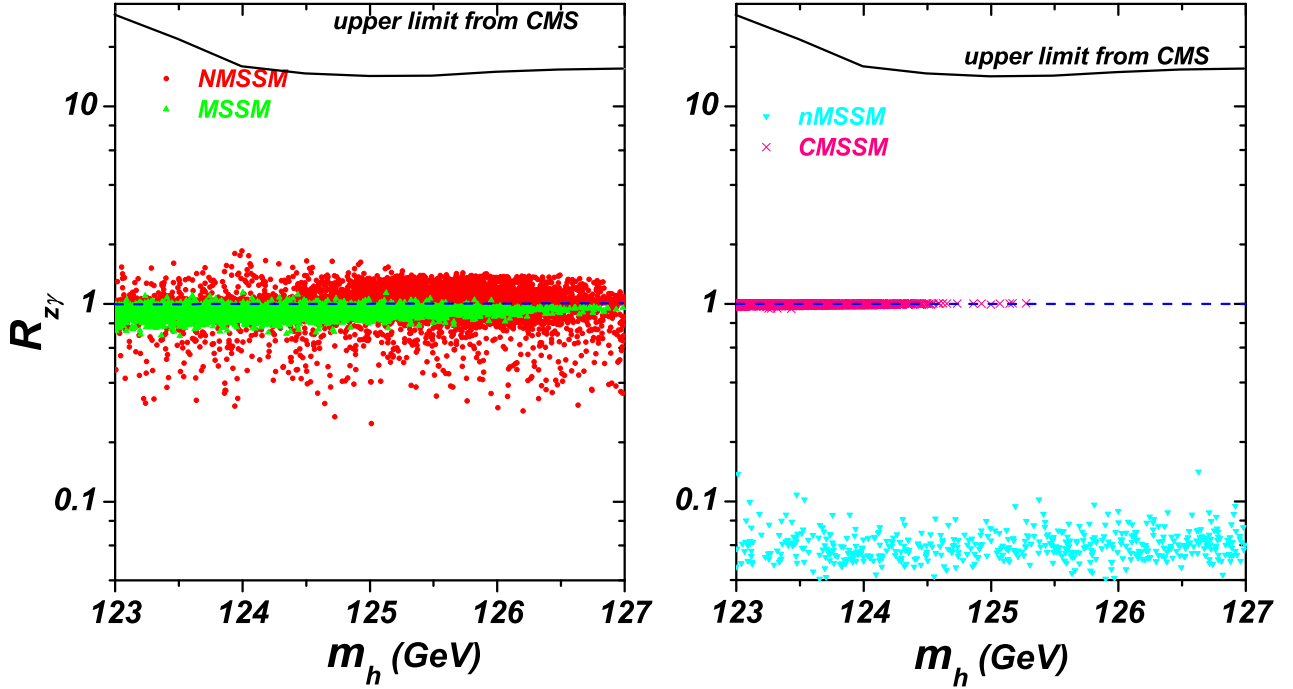}
\vspace{-1.4cm}
\caption{The scatter plots of the surviving samples in the four models, projected on the $R_{Z\gamma}-m_h$ plane.
Here $R_{Z\gamma}=\sigma(pp\to h\to Z\g)/\sigma_{\rm SM}(pp\to h\to Z\g)$ denotes the normalized $Z \gamma$ signal rate in
the SM-like Higgs boson production at the LHC, and the black line represents its upper
limit set by the CMS collaboration \cite{zg-exp}.}
\label{fig1}
\end{figure}
In Fig.\ref{fig1} we project the surviving samples on the plane of the $Z\gamma$ signal
rate at the LHC versus the SM-like Higgs boson mass in the four SUSY models. We also
show the CMS bound on the  rate in the figure \cite{zg-exp}.
From the left panel we see that compared with its SM prediction, the $Z\gamma$  rate
in the MSSM and the NMSSM can be either enhanced or suppressed with the maximal deviation reaching
$20\%$ and $60\%$ respectively. In contrast, as shown in the right panel, the $Z\gamma$ rate is always
slightly suppressed (less than $5\%$) in the CMSSM and severely suppressed
(more than $90\%$) in the nMSSM. We checked that for the CMSSM and the MSSM, the suppression is
mainly due to the increase of $h \to b \bar{b}$ partial width \cite{diphoton-SUSY-2,diphoton-SUSY-3}.
For the nMSSM, however, it is due to the open up of new decays $h\to \chi^0 \chi^0, a_1a_1$
($\chi^0$ and $a_1$ denote the dark matter and lightest CP-odd higgs boson respectively), which
significantly enlarges the total width of the Higgs boson and leads to
severe suppression for all visible decay channels.

For the samples shown in Fig.\ref{fig1}, we also compare their predictions on the
$\gamma\gamma$ and $ZZ^\ast$ signal rates of the Higgs boson with the corresponding
experimental data. We find that for most of the samples in the MSSM and the NMSSM
and for all the samples in the CMSSM, their theoretical predictions on the
$\gamma\gamma$ and $ZZ^\ast$ rates agree with the corresponding data at $3 \sigma$ level,
while for the samples in the nMSSM, their predictions always lie outside the $3\sigma$
regions (see Fig.1 and Fig.2 in \cite{diphoton-SUSY-4}). Moreover, we checked that the
branching ratio of $h \to b \bar{b}$ in the MSSM, NMSSM and CMSSM varies in the
ranges [$57\%$, $69\%$], [$32\%$, $67\%$] and [$60\%$, $63\%$], respectively
(in the SM its value is about $57\%$ for $m_h \simeq 125.5 {\rm GeV}$),
and the signal strength for the process $p p \to V h \to V b\bar{b}$ normalized by
its SM value varies from 0.97 to 1.12, 0.55 to 1.05 and 1.00 to 1.02, respectively.
Considering that so far the only way
to detect the $h \to b \bar{b}$ decay at the LHC is through the $Vh$ associated production,
whose signal strength $\mu_{Vb\bar{b}}$ is $-0.4\pm 1.0$ from ATLAS result
\cite{ATLAS-bb} and $1.0\pm 0.49$ from CMS result \cite{CMS-bb},
one can conclude that such alterations of $b\bar{b}$ signal rates
are allowed by the current experimental data at $2\sigma$ level.

\begin{figure}[t]
\includegraphics[width=16cm]{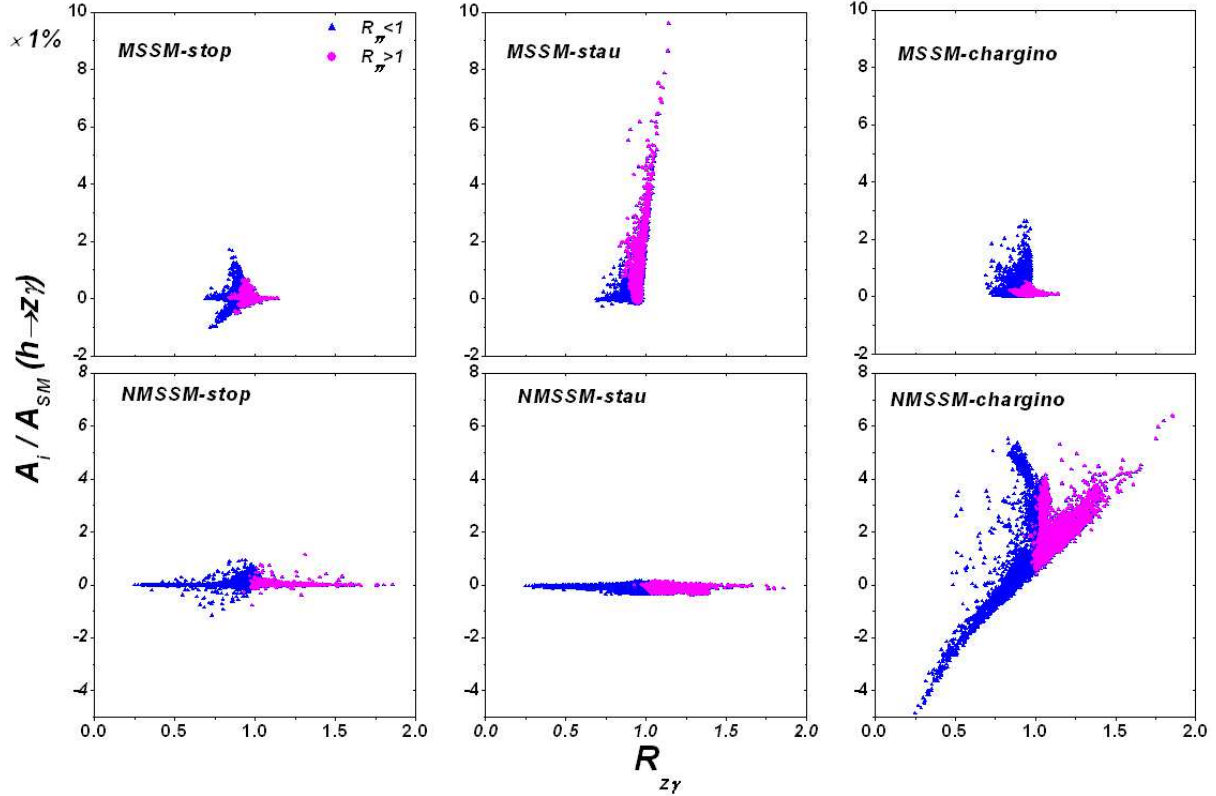}
\vspace{-0.8cm}
\caption{Same as Fig.1, but showing the normalized sparticle
contributions to the amplitude of $h \to Z\g$ in the MSSM and NMSSM. The magenta bullets and
blue triangles represent the samples with $R_{\g\g}>1$ and $R_{\g\g}<1$ respectively.} \label{fig2}
\end{figure}
Next we focus on the MSSM and the NMSSM. In Fig.2 we exhibit the contributions of different sparticles to
the amplitude of the $h \to Z\g$ decay. Since the sbottoms and charged Higgs bosons have
little effect on the amplitude, we do not show their contributions.
This figure shows following features:
\begin{itemize}
\item[(1)] In the MSSM, the potentially largest contribution comes from the stau loops, which
can alter the SM amplitude by about $10\%$. In contrast, the stop contribution is small, which
usually changes the amplitude by less than $3\%$. This feature can be well understood
by the formulae listed in Appendix.
Explicitly speaking, in order to get a significant sfermion contribution, one necessary
condition is $Y_{hLR}\sin 2 \theta_{\tilde{f}}/m_{\tilde{f}_1}^2 $ should be as large as possible,
where $Y_{hLR}$ denotes the chiral flipping coupling of Higgs to sfermions,
$\theta_{\tilde{f}}$ is the chiral mixing angle and $m_{\tilde{f}_1}$ represents the lighter
sfermion mass. As far as the stop sector is concerned, a relatively light
$\tilde{t}_1$ is always accompanied by a heavy $\tilde{t}_2$ in order to predict
$m_h \simeq 125 {\rm GeV}$. Although $A_t$ in this case may be very large, the chiral mixing angle $\theta_{\tilde{t}}$
is usually small and consequently, the stop contribution can never get significantly enhanced.
In the stau sector, however,
both the parameters $M_{L3}$ and $M_{E3}$ are unlimited, and one can choose
light staus and an appropriate $\theta_{\tilde{\tau}}$ to maximize the contribution.
In this process, the value of $\mu \tan \beta$
and the splitting between $M_{L3}$ and $M_{E3}$ play an important
role. It is worth noting that a light stau with mass close to dark matter may co-annihilate with the dark matter,
which is helpful to avoid the overabundance of the dark matter in today's universe\cite{light-stau-coannihilation-Ellis-9810360,light-stau-coannihilation-Ellis-9905481}.

\item[(2)]
In the MSSM, the chargino contribution is small and can only reach
$3\%$ and $0.5\%$ for $R_{\gamma \gamma}<1$ case and $R_{\gamma \gamma} > 1$ case respectively.
The reason is that in the MSSM, the $h \chi^+_1 \chi^-_1$ coupling is induced by the
$H^0_i \bar{\tilde{H}} \tilde{W}$ interaction ($i=u,d$ and $\tilde{H}$ and $\tilde{W}$ denote
Higgsino and Wino respectively), and this coupling strength is maximized when both the Higgsino
and Wino components of $\chi^\pm_1$ are sizable. We checked that, for most of the $R_{\gamma \gamma} > 1$
samples,  $M_2 \le 700 {\rm GeV}$ and $\mu > 800 {\rm GeV}$ (a large $\mu$ is needed for the
stau contribution to enhance the diphoton rate \cite{diphoton-SUSY-3}) so that $\chi^\pm_1$
is basically Wino-like. Consequently, its coupling to the Higgs boson is weak.

\item[(3)] In the NMSSM, the largest SUSY contribution to the $Z\gamma$ decay comes
from the chargino loops with the correction reaching $6\%$ in optimal case, while the magnitude
of the stau contribution is always smaller than $1\%$. This is because in the NMSSM with a large $\lambda$,
$\mu$ is preferred to vary from 100 GeV to 250 GeV and $\tan \beta$ is usually smaller than 10 \cite{fine-tuning-NMSSM-Ellwanger:2011mu}. As a result, the $h \bar{\chi}_1^\pm \chi_1^\pm$ coupling
is relatively large, while the $h\tilde{\tau}_L^\ast \tilde{\tau}_R$ coupling
can not be pushed up by the moderate $\mu \tan\beta$. Due to the singlet component of $h$ in the NMSSM,
the $hb\bar{b}$ coupling can be greatly suppressed (reaching $40\%$ by our results) so that the total width of $h$ is reduced by about $50\%$ in extreme case\cite{diphoton-SUSY-4}. Consequently, even when the SUSY  contributions to the decay width are small,
$R_{Z\gamma}$ can still be quite large.
As we mentioned before,
such a suppressed $hb\bar{b}$ coupling can reduce the normalized strength of the
process $p p \to V h \to V b\bar{b}$ down to 0.55, which, however, is
still compatible with the current LHC data due to the large uncertainty
of the measured signal strength ($\mu_{Vb\bar{b}} = -0.4 \pm 1.0$ from ATLAS result \cite{ATLAS-bb}
and $\mu_{Vb\bar{b}} = 1.0 \pm 0.49$ from CMS result \cite{CMS-bb}).
\end{itemize}

\begin{figure}[t]
\includegraphics[width=17.5cm]{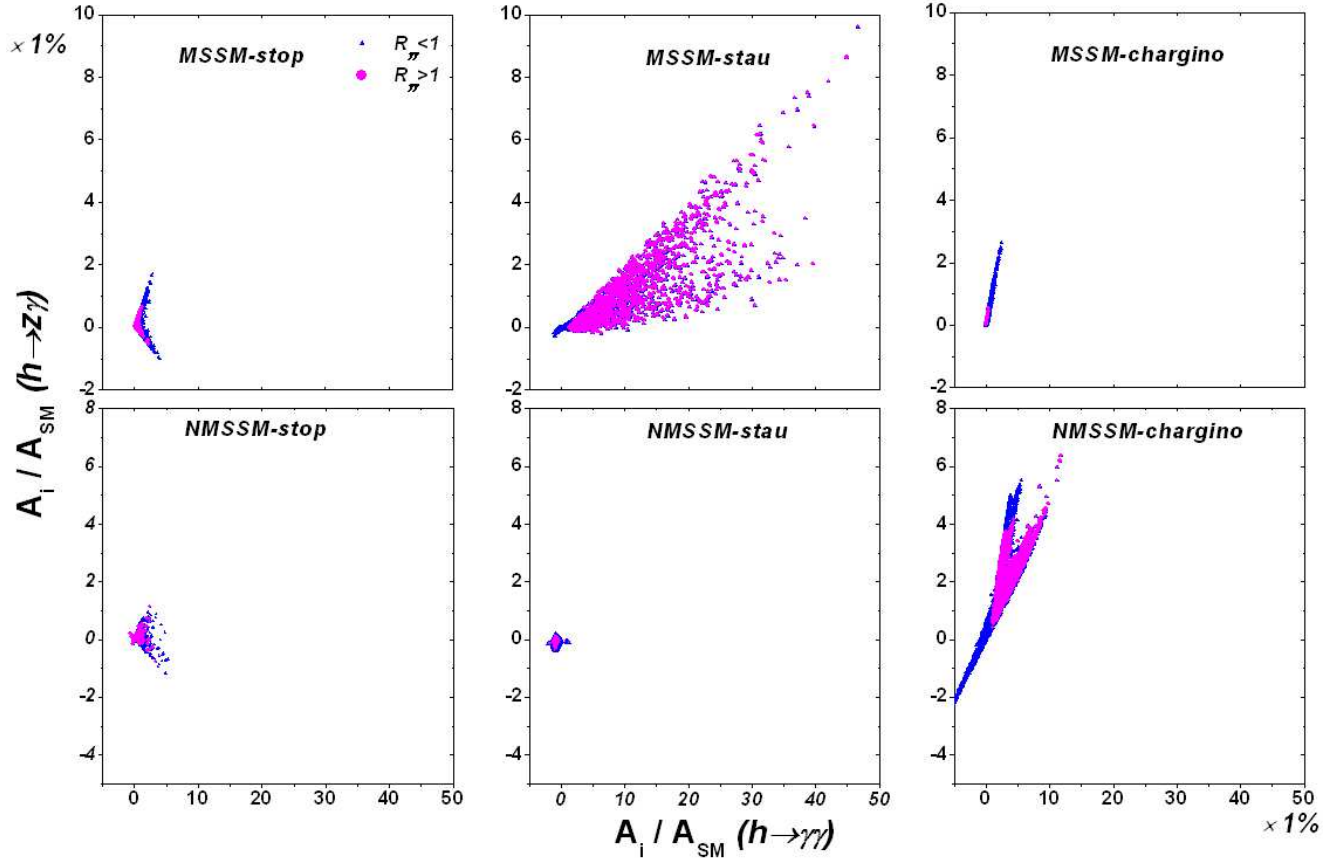}
\vspace{-1.5cm}
\caption{Same as Fig.2, but showing the correlation of
SUSY particle contribution to $h\to Z\g$ channel with that to $h\to \g\g$ channel.}
\label{fig3}
\end{figure}
In Fig.\ref{fig3}, we show the correlation of the amplitudes for $h\to Z \gamma$ and
$h \to \gamma \gamma$. This figure indicates that in both the MSSM and the NMSSM, the top squark contribution to
the amplitude of $h\to Z \gamma$ correlates roughly in a linear way with that of $h\to \gamma \gamma$, and
so is the chargino contribution. Fig.\ref{fig3} also indicates that the correlation is spoiled for the stau
contribution in the MSSM. We checked that this is because $\theta_{\tau}$ in the MSSM can vary over
a broad range and the dependence of the two amplitudes on  $\theta_{\tau}$ are quite different.
For example, in the case $M_{L3}\simeq M_{E3}$, $\theta_\tau \simeq \pi/4$ and both $Z \tilde{\tau}_i^\ast \tilde{\tau}_i$
and $h \tilde{\tau}_1^\ast \tilde{\tau}_2$ couplings approach zero by accidental cancelation.
As a result, the stau contribution to the decay $h \to Z \gamma$ is suppressed. In contrast, the contribution
to the decay $h\to \gamma \gamma$ is maximized since it is proportional to $\sin 2 \theta_{\tau}$.
On the other hand, if $|M_{L3}^2-M_{E3}^2| \gg m_\tau \mu \tan \beta$ so that $\theta_\tau \to 0$,
the contributions are suppressed for both the decays because the dominant contribution to
$Z \tilde{\tau}_1^\ast \tilde{\tau}_2$ coupling and that to $h \tilde{\tau}_i^\ast \tilde{\tau}_i$
coupling are both proportional to $\sin 2 \theta_{\tau}$.

\begin{figure}[t]
\includegraphics[width=15cm]{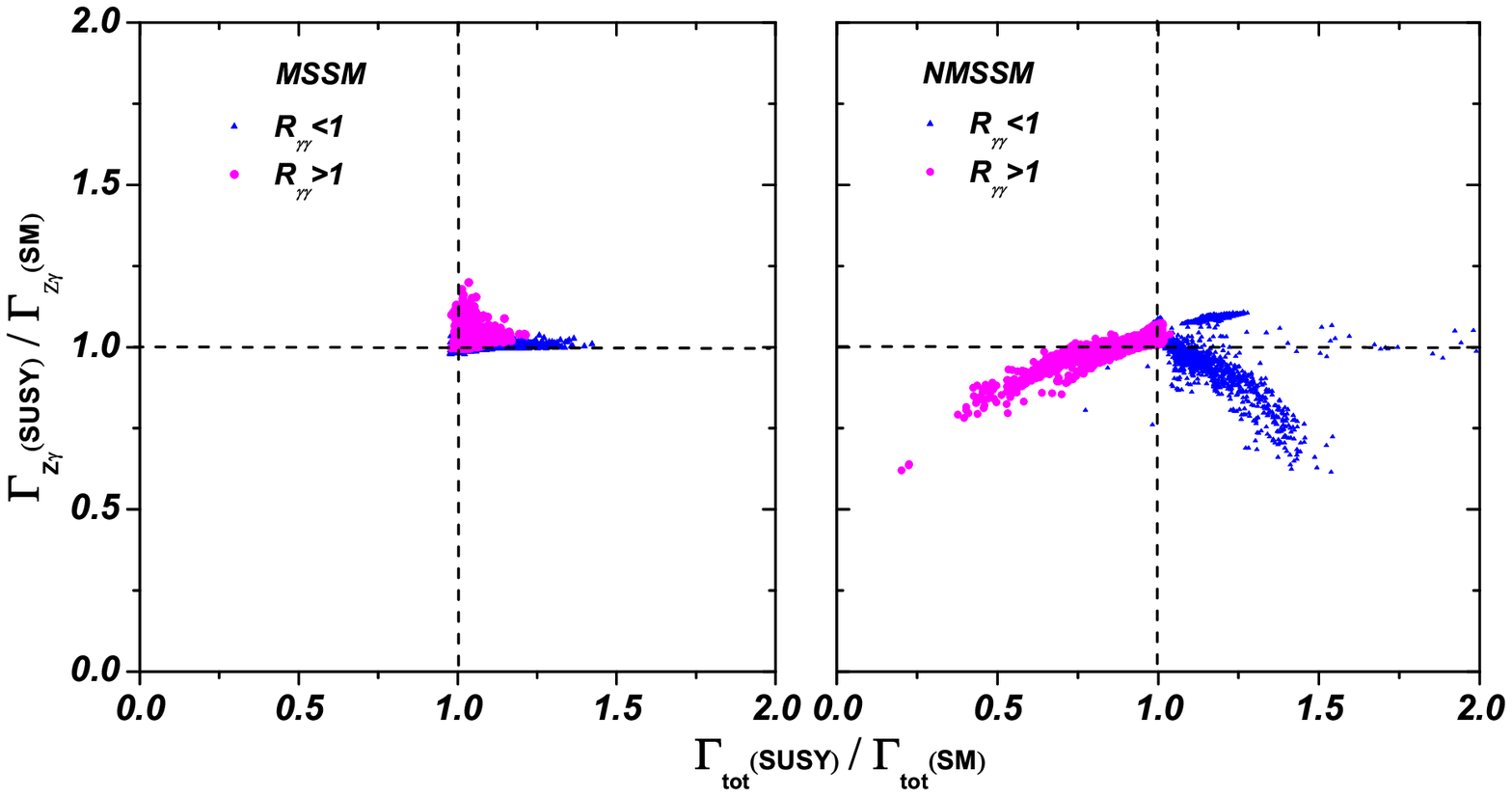}
\vspace{-0.3cm}
\caption{Same as Fig.2, but projected on
the plane of $\G^{\rm SUSY}_{Z\g}/\G^{\rm SM}_{Z\g}$ versus
$\G^{\rm SUSY}_{\rm total}/\G^{\rm SM}_{\rm total}$. }
\label{fig4}
\end{figure}
\begin{figure}[t]
\includegraphics[width=17.5cm]{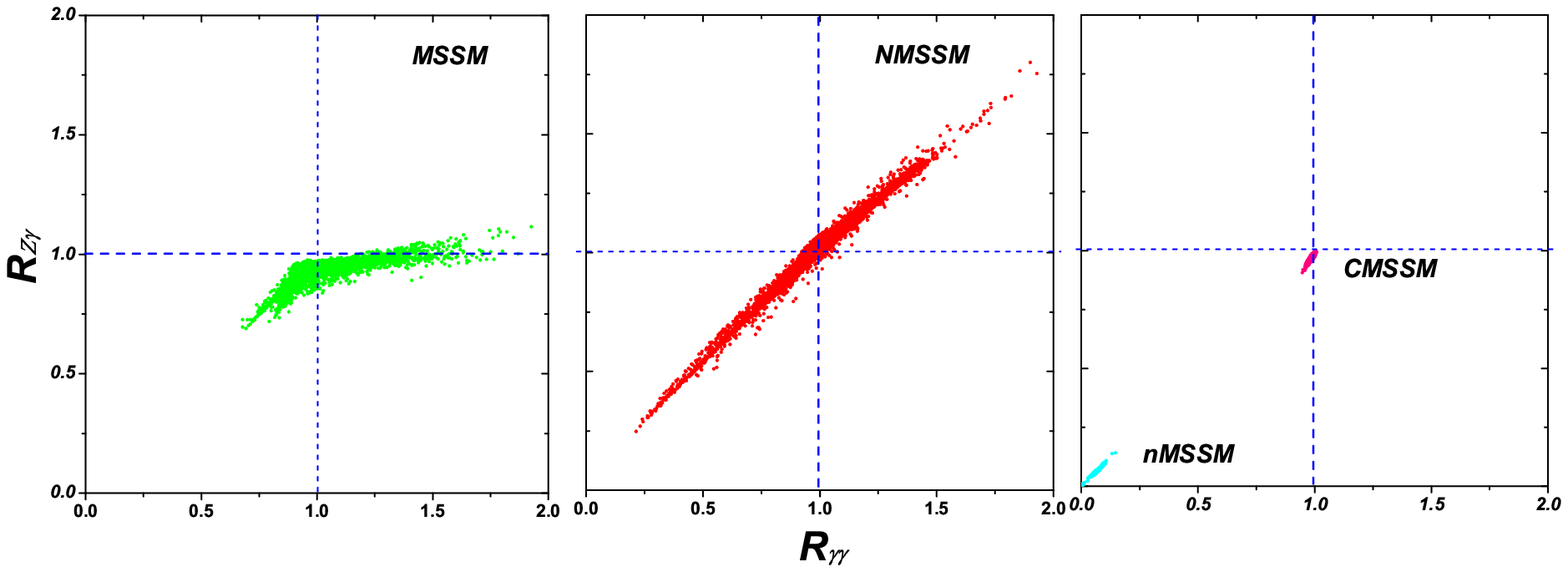}
\caption{Same as Fig.1, but showing the correlation between
$R_{Z\g}=\sigma(pp\to h\to Z\g)/\sigma_{\rm SM}(pp\to h\to Z\g)$
and $R_{\g\g}=\sigma(pp\to h\to \g\g)/\sigma_{\rm SM}(pp\to h\to \g\g)$
in the four models.}
\label{fig5}
\end{figure}

Considering  $R_{Z\gamma}$ is mainly determined by the partial
width of $h \to Z\g$ and the total width of the SM-like Higgs boson,
we present in Fig.\ref{fig4} the ratio of $\G^{\rm SUSY}_{Z\g}/\G^{\rm SM}_{Z\g}$ versus the ratio of
$\G^{\rm SUSY}_{\rm total}/\G^{\rm SM}_{\rm total}$ for the two models.
The left panel indicates that for almost all MSSM samples  the $Z\g$ partial width and the total
width of the SM-like Higgs boson is larger than the corresponding SM predictions. These features
originate from the constructive contributions of the SUSY particles to $h \to Z \gamma$ and
the enhanced  width of $h \to b\bar{b}$ respectively. Interestingly, the largest increase of
$\Gamma_{Z\gamma}$ occurs when $\Gamma_{\rm total}^{\rm SUSY} \simeq \Gamma_{\rm total}^{\rm SM}$.
The right panel indicates that, in order to enhance the $Z \gamma$ signal in the NMSSM with
a large $\lambda$ (note that in this model, $R_{Z\gamma}$ correlates roughly in a linear way
with $R_{\gamma \gamma}$, see Fig.\ref{fig5}), the SM-like Higgs boson tends to have sizable singlet
component to suppress the total width. In this case, $\Gamma_{Z\gamma}$ is suppressed too, but
we have $\Gamma_{Z\gamma}^{\rm SUSY}/\Gamma_{Z\gamma}^{\rm SM} >
\Gamma_{\rm total}^{\rm SUSY}/\Gamma_{\rm total}^{\rm SM}$.  Moreover, as mentioned
before, $\Gamma_{Z\g}$ can be slightly enhanced by the chargino contribution.

Now we investigate the correlation of the $Z\gamma$ rate with the $\gamma \gamma$ rate
in different SUSY models, which is shown in Fig.\ref{fig5}.
From this figure we have following conclusions:
\begin{itemize}
\item[(a)] In the MSSM, although the partial width of $h \to Z \gamma$ can be enhanced
by $20\%$ (see Fig.\ref{fig4}), due to the increase of the Higgs total width and also
the suppression of the $h g g$ coupling \cite{diphoton-SUSY-3}, the maximal value of
$R_{Z\gamma}$ is only 1.1 (in comparison, $R_{\gamma \gamma}$ may be as large as 2.),
and only when $R_{\gamma \gamma} \gtrsim 1.25$ can $R_{Z\gamma} > 1$ be possible.
Among the sparticle contributions to $R_{Z\gamma}$ and $R_{\gamma\gamma}$, the stau
loops play the dominant role. The difference between the two signals comes from
their dependence on $\theta_{\tau}$, i.e.  $\theta_{\tau} \simeq \pi/4$ $R_{\gamma \gamma}$
is maximized while $R_{Z\gamma}$ is suppressed.
Our numerical results also indicate that,
for the surviving samples of the MSSM, the branching ratio of the invisible decay
$h\to\chi^0 \chi^0$ is usually smaller than $6\%$ and $3\%$ for $R_{\gamma \gamma}<1$
and $R_{\gamma \gamma}>1$ case, respectively, and the branching ratio of
$h \to b \bar{b}$ varies from $57\%$ to $69\%$.
\item[(b)] In the NMSSM with a large $\lambda$, the sparticle corrections to
the amplitudes of $h\to Z \gamma$ and $h\to \gamma \gamma$ are usually below $10\%$, and the
main mechanism to alter $R_{Z\gamma}$ and $R_{\gamma \gamma}$
is through the suppression of the $hb\bar{b}$ and $hW^+W^-$ couplings by the singlet component of $h$.
As a result, $R_{Z\gamma}$ and $R_{\gamma \gamma}$ are highly correlated and
both of them vary from 0.2 to 2.
We checked that the branching ratios of
the exotic decays $h\to \chi^0 \chi^0, a_1a_1$ may reach $22\%$ and $45\%$, respectively
(these extreme cases correspond to some of the squared points in Fig.5 of
\cite{diphoton-SUSY-4}), and the branching ratio of the decay $h \to b \bar{b}$
varies in a large range, from $32\%$ to $67\%$.
\item[(c)] In the CMSSM and nMSSM, $R_{Z\gamma}$ and $R_{\gamma \gamma}$  are slightly
and strongly suppressed respectively. As discussed before, in the CMSSM the
suppression is due to the increase of $h\to b \bar{b}$ partial width,
while in nMSSM it is due to the open up of the
exotic decay channels $h\to \chi^0 \chi^0, a_1 a_1$.
\end{itemize}

Note that the open up of the exotic decays $h\to \chi^0 \chi^0, a_1a_1$
will generally lead to the suppression of visible signal rates such as
$R_{\gamma \gamma}$ and $R_{Z Z^\ast}$,
and as analyzed in \cite{Giardino_large_exotic_decay_analysis},
the latest Higgs data require that the
total branching ratio of the exotic decays should be less than $28\%$ at $95\%$ C.L..
This conclusion again indicates that the nMSSM and also some samples of the NMSSM
are disfavored by the current Higgs data (about this conclusion, one may also see
the squared points in Fig.5 of \cite{diphoton-SUSY-4}).
For the decay $h \to a_1 a_1$ in the NMSSM,
we checked that $m_{a_1}$ varies from about $20 {\rm GeV}$ to $60 {\rm GeV}$, and $a_1$ mainly decays
to $b\bar{b}$ (with a branching ratio at about $90\%$)
and $\tau\bar{\tau}$ (with the branching ratio at about $9\%$). So in this case, the decay product of the Higgs boson
is four $b$-jets, or four $\tau$ leptons or two $b$-jets plus two $\tau$ leptons, which is an interesting but
challenging signal in Higgs search at the LHC\cite{hard-scenario}. Since $m_{a_1}$ is usually heavier than $\Upsilon$,
the constraint from the decay $\Upsilon\to a_1\gamma $ \cite{CLEO-Upsilon-A-gamma} is irrelevant here.
Also note that generally speaking, the constrained model such as CMSSM
tends to predict strong correlations among observables due to the unified nature of its parameters.
This is clearly shown in Fig.\ref{fig5} for the CMSSM in comparison with the other three models.

\begin{figure}[t]
\includegraphics[width=17.5cm]{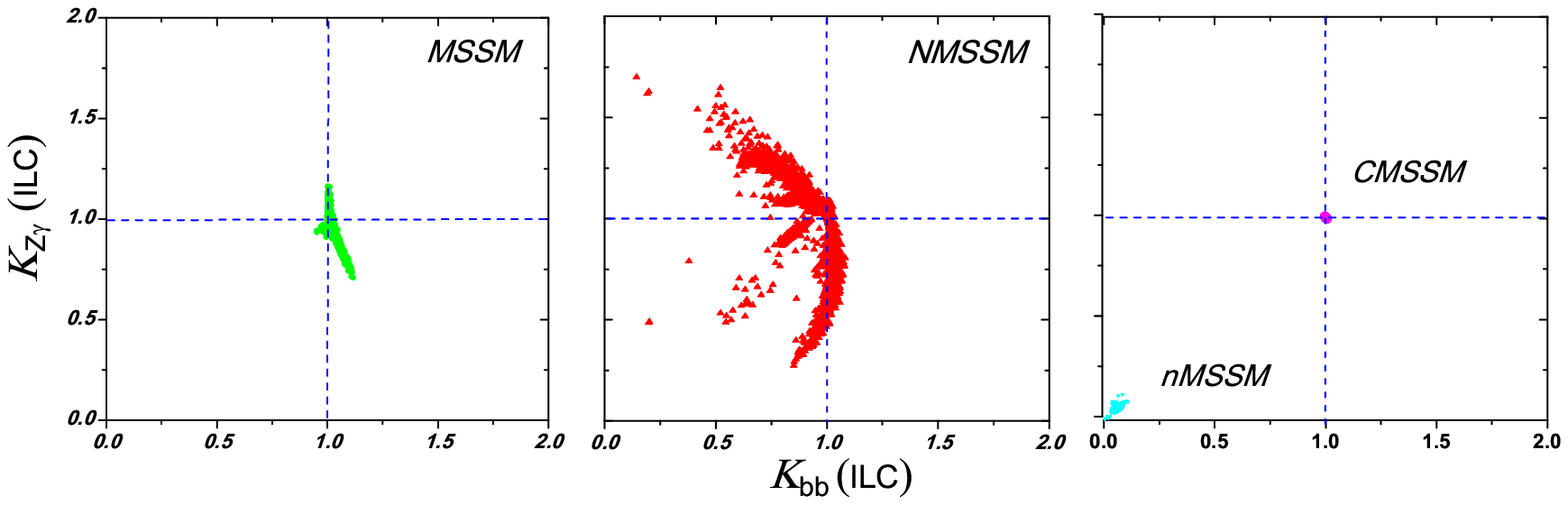}
\caption{Same as Fig.1, but showing the correlation between
${\cal{K}}_{Z\g}=\sigma(e^+e^-\to Z h\to Z Z\gamma)/\sigma_{\rm SM}(e^+e^-\to Z h\to Z Z\gamma)$
and ${\cal{K}}_{b\bar{b}}=\sigma(e^+e^-\to Z h\to Z b\bar{b})/\sigma_{\rm SM}(e^+e^-\to Z h\to Z b\bar{b})$
in the four models.}
\label{fig6}
\end{figure}
Since the $e^+e^-$ collider at $\sqrt{s}\sim250 {\rm GeV}$
provides a clean environment to detect the decays $h\to Z\gamma$ and $h\to b\bar{b}$,
we also investigate their rates at the ILC defined in Eq.(\ref{ILC1}) and Eq.(\ref{ILC2}).
The corresponding results are shown in Fig.\ref{fig6}. This figure exhibits following features
\begin{itemize}
\item[(a)] In the MSSM, a suppressed $Z\gamma$ signal (compared with its SM prediction) tends to correspond to an
enhanced $b\bar{b}$ signal, and an enhanced $Z\gamma$ signal requires the $b\bar{b}$ signal rates to be roughly
at its SM prediction. In any case, the enhancement factor for the two signals are less than 1.2.  Note that there
exist a few cases where the $b\bar{b}$ signal rate is slightly suppressed.
\item[(b)] In the NMSSM with a large $\lambda$, the normalized $b\bar{b}$ signal rate is less than 1.1, and in some
cases it may be significantly suppressed. In contrast, the $Z\gamma$ signal rate can be either greatly enhanced or
severely suppressed. In the enhancement case, the $b\bar{b}$ signal rate is usually less than its SM prediction, and
the greater enhancement corresponds to the stronger suppression.
\item[(c)] In the CMSSM, both the signal rates are roughly equal to their SM predictions. In the nMSSM, however, both the rates are strongly suppressed.
\end{itemize}
Moreover, we checked that the $\gamma \gamma$ signal rate at the ILC has similar dependence on the $b\bar{b}$ rate for the
four models.

\section{conclusion}
In this work, we investigate the rare decay of the SM-like Higgs boson, $h\to Z\g$, and study its correlation with
$h\to \g\g$ in the MSSM, the NMSSM, the nMSSM and the CMSSM. We perform a scan over the parameter space of each
model by considering various experimental constraints and present our results on various planes.
We have  following observations:
\begin{itemize}
\item[(i)] In the SUSY models, the sparticle correction to the rare decay  $h\to Z \gamma$ is usually several times smaller
 than that to $h \to \gamma \gamma$.
\item[(ii)] In the MSSM, the net SUSY contribution to the amplitude of $h\to Z\g$ is constructive with the corresponding
SM amplitude and can enhance the SM prediction by at most $10\%$. As a result, the $Z\gamma$ signal rates at
the LHC and the ILC can be enhanced by $20\%$ at most. As a comparison, the $\gamma \gamma$ rate can be enhanced by
a factor of 2 due to the large stau contributions.
\item[(iii)] In the CMSSM, due to the slightly enhanced total width of the SM-like Higgs bson, the $Z \gamma$ signal
 rates at the LHC and the ILC are both slightly below their SM predictions, and so is the $\g\g$ signal.
\item[(iv)] In the NMSSM with a large $\lambda$, the SUSY corrections to the amplitudes for the
decays $h \to Z \gamma$ and $ h \to \gamma \gamma$ are at most $10\%$, and to get significant deviation of
the two rates from their SM values, the total width of the SM-like Higgs boson must be
moderately suppressed by the singlet component of $h$. In this model, the two rates are highly correlated and vary from 0.2 to 2.
\item[(v)] In the nMSSM, the signal rates of $h\to Z\g$ and  $h\to \g\g$ are both
greatly suppressed due to the open up of the exotic decay $h \to \chi^0\chi^0, a_1a_1$.
\end{itemize}

Finally, we note that some strategies have been proposed in the literature to discriminate the considered models, e.g.,
via the correlations between the Higgs couplings \cite{diphoton-SUSY-4},
via the enhanced Higgs pair productions at the LHC \cite{higgs-pair-lhc}
and the ILC \cite{higgs-pair-ilc}, or via the direct dark matter detection
\cite{Cao-2010}.
Compared with these existing strategies,
the loop-induced $Z\gamma$ and $\gamma \gamma$ decay modes of the Higgs
boson seem to be more sensitive to the nature of the models
(some non-SUSY models predict rather different correlation behavior \cite{new-scalar-zg}).
So the correlation between  $Z\gamma$ and $\gamma \gamma$ rates
analyzed in this work may play a complementary role to discriminate
new physics models in the future.

{\em Note added:~~}
After we finished this work, both the ATLAS and CMS
collaborations updated their Higgs search results \cite{1303-a-2ph,1303-c-2ph}.
Among these new results, the CMS data on the diphoton signal rate is changed
significantly \cite{1303-c-2ph}.
Since in our analysis we did not use the diphoton data as a constraint (instead we just
displayed the predictions for the diphoton signal rate in different SUSY models),
our results and conclusions are not affected by the new data.

\section*{Acknowledgement}
We thank Kun Yao and Jingya Zhu for helpful discussions. This work
was supported in part by the National Natural Science Foundation of
China (NNSFC) under grant No. 10775039, 11075045,
11275245, 10821504 and 11135003, by the Project
of Knowledge Innovation Program (PKIP) of Chinese Academy of
Sciences under grant No. KJCX2.YW.W10.

\section*{Appendix}

In SUSY, the decays $h\to Z\gamma $ and $h \to \gamma\gamma$ get new contributions from the loops mediated by
charged Higgs bosons, sfermions (including stops, sbottoms and staus)  and charginos, and as a result, the formula of
$\Gamma_{Z\gamma}$ and $\Gamma_{\gamma\gamma}$ are modified by
\begin{eqnarray}
\Gamma (h \to Z\gamma) &=& \frac{G^2_F m_W^2\,\alpha\,m_h^{3}}
{64\,\pi^{4}} \left(1-\frac{m_Z^2}{m_h^2} \right)^3 \bigg| {\cal
A}_{W}^{Z\g}+ {\cal A}_{t}^{Z\g}+{\cal
A}_{H^{\pm}}^{Z\g}+{\cal A}_{\tilde{f_i}}^{Z\g}+{\cal
A}_{\chi^{\pm}_i}^{Z\g} \bigg|^2\\
\Gamma(h \rightarrow\gamma\gamma) &=& \frac{G_F\alpha^2
m_h^3}{128\sqrt{2}\pi^3}\bigg|{\cal A}_{W}^{\g\g}+ {\cal
A}_{t}^{\g\g}+{\cal
A}_{H^{\pm}}^{\g\g}+{\cal A}_{\tilde{f_i}}^{\g\g}+{\cal A}_{\chi^{\pm}_i}^{\g\g}\bigg|^2.
\end{eqnarray}
The expressions of $A_i^{\gamma \gamma}$ are relatively simple and are given by
\begin{eqnarray}
&&{\cal A}_{W}^{\g\g}=g_{hVV} A_1(\tau_W),
~{\cal A}_{t}^{\g\g}=g_{ht\bar{t}} N_c Q_t^2 A_{1/2}(\tau_t),
~ {\cal A}_{\tilde{f_i}}^{\g\g}=\sum_{i}\frac{ g_{h\tilde f_i \tilde f_i}
}{m_{\tilde{f}_i}^2} N_c Q_{\tilde f_i}^2 A_0(\tau_{ {\tilde f}_i}),\nonumber\\
&&{\cal A}_{H^{\pm}}^{\g\g}=\frac{m_Z^2 g_{h H^+ H^-}}{2
M_{H^\pm}^2} A_0(\tau_{H^\pm}),
~{\cal A}_{\chi^{\pm}_i}^{\g\g}=\sum_{i} \frac{2 m_W}{ m_{\chi_i^\pm}}
g_{h\chi_i^+\chi_i^-} A_{1/2} (\tau_{\chi_i^\pm}),
\end{eqnarray}
where $\tau_i=4m_i^2/m_h^2$, $g_{hXY}$ denotes the Higgs coupling with particles $XY$ and $A_0, A_{1/2}$ and $A_1$ are loop functions
with scalars, fermions and gauge bosons running in the loop. The explicit expressions of $g_{hXY}$ and $A$ functions are given by
\begin{eqnarray}
&&g_{hVV} = S_{h1}\sin\beta+S_{h2}\cos\beta, \\
&&g_{ht\bar{t}}= S_{h1}/\sin\beta, \\
&&g_{h\tilde f_1 \tilde f_1} = \frac{-1}{2(\sqrt{2}G_F)^{1/2}}\bigg(
 \mathrm{Y}_{hLL}\cos^2\theta_{\tilde f}+\mathrm{Y}_{hRR}\sin^2\theta_{\tilde f}+\mathrm{Y}_{hLR}\sin2\theta_{\tilde f} \bigg), \\
&&g_{h\tilde f_2 \tilde f_2} = \frac{-1}{2(\sqrt{2}G_F)^{1/2}}\bigg(
 \mathrm{Y}_{hLL}\sin^2\theta_{\tilde f}+\mathrm{Y}_{hRR}\cos^2\theta_{\tilde f}-\mathrm{Y}_{hLR}\sin2\theta_{\tilde f} \bigg), \\
&&g_{h H^+ H^-} =
 \frac{\lambda^2}{\sqrt{2}} \bigg( v_s(\Pi_{h3}^{11}+\Pi_{h3}^{22})-v_u\Pi_{h2}^{12}  -v_d\Pi_{h1}^{12}   \bigg)
 +\sqrt{2}\lambda\kappa v_s \Pi_{h3}^{12}
 +\frac{\lambda}{\sqrt{2}}A_\lambda \Pi_{h3}^{12}\nonumber\\
&&\quad \quad \quad \quad +\frac{g_1^2}{4\sqrt{2}}\bigg( v_u(\Pi_{h1}^{11}-\Pi_{h1}^{22}) +v_d(\Pi_{h2}^{22}-\Pi_{h2}^{11})  \bigg)\nonumber\\
&& \quad \quad \quad \quad +\frac{g_2^2}{4\sqrt{2}}\bigg( v_u(\Pi_{h1}^{11}+\Pi_{h1}^{22}+2\Pi_{h2}^{12})
 +v_d(\Pi_{h2}^{11}+\Pi_{h2}^{22}+2\Pi_{h1}^{12})  \bigg),\nonumber\\
&&\Pi_{hi}^{jk}=2S_{hi} C_j C_k, \quad C_1=\cos\beta,\quad C_2=\sin\beta,\nonumber\\
&&v_u= \frac{1}{(\sqrt{2}G_F)^{1/2}}C_2, \quad v_d=\frac{1}{(\sqrt{2}G_F)^{1/2}} C_1, \quad v_s=\mu/\lambda,\\
&&g_{h \chi^+_i \chi^-_j}^L =\frac{1}{\sqrt{2}}(S_{h1}U_{i1}V_{j2}+S_{h2}U_{i2}V_{j1}),\quad g_{h \chi^+_i \chi^-_j}^R=\frac{1}{\sqrt{2}}(S_{h1}U_{j1}V_{i2}+S_{h2}U_{j2}V_{i1}),\\
&&A_0(x) =-x^2 \left[x^{-1}-f(x^{-1})\right],\\
&&A_{1/2}(x) = 2  \, x^2 \left[x^{-1}+ (x^{-1}-1)f(x^{-1})\right],\\
&&A_1(x)=-x^2\left[2x^{-2}+3x^{-1}+3(2x^{-1}-1)f(x^{-1})\right],
\end{eqnarray}
where S is the $2\times 2$ ($3\times 3$) rotation matrix of MSSM (NMSSM) higgs mass matrix under the basis $(H^0_u, H^0_d, S)$, h in $S_{h1}$ denotes
the row index of the SM-like Higgs, $Y_{hXY}$ denotes the
SM-like higgs coupling to sfermion interaction states, $U,V$ denote the rotation matrices of the chargino mass matrix,
and $f(x)$ is defined by $f(x) = \arcsin^2 \sqrt{x}$.

As for $A_i^{Z\gamma}$, due to $m_Z \neq 0$ and the existence of $Z X Y$ ($X\neq Y$) couplings, their expressions
are rather complex
\begin{eqnarray}
&&{\cal A}_{W}^{Z\g}=g_{hVV} c_w A_1 (\tau_W,\lambda_W),~{\cal
A}_{t}^{Z\g} = g_{ht\bar{t}} N_c \, Q_t \frac{\hat{v}_t }{c_w} \,
A_{1/2} (\tau_t,\lambda_t),\nonumber \\
&&{\cal A}_{H^{\pm}}^{Z\g}=-\frac{m_Z^2 g_{ h H^+ H^-} } {2\, m_{H^\pm}^2
}v_{H^\pm} A_0 (\tau_{H^\pm},\lambda_{H^\pm}),\nonumber \\
&&{\cal A}_{\tilde{f_i}}^{Z\g}=-\sum_{\tilde f_i} \frac{2\, g_{h \tilde f_i
\tilde f_i}}{ m_{\tilde{f}_i}^2} N_c Q_{\tilde f_i} v_{\tilde f_i}
A_0 (\tau_{ {\tilde f}_i},\lambda_{ {\tilde f}_i})
- {\cal A}_{\tilde{f}_1 \tilde{f}_2}^{Z\g},\nonumber \\
&&{\cal A}_{\tilde{f}_1 \tilde{f}_2}^{Z\g}=\frac{2\, g_{h \tilde f_1
\tilde f_2}}{ m_{\tilde{f}_1} m_{\tilde{f}_2}} N_c Q_{\tilde f} v_{\tilde f_{12}}(A^{(1)}_0 +A^{(2)}_0 ),\nonumber \\
&&{\cal A}_{\chi^{\pm}_{i}}^{Z\g}=\sum\limits_{\chi^{\pm}_{i}; m,n=L,R} \frac{2 m_W}{
m_{\chi_i^\pm}} g_{h\chi^+_i \chi^-_i}^m g_{Z \chi^+_i \chi^-_i}^n
A_{1/2}(\tau_{\chi^\pm_i}, \lambda_{\chi^\pm_i})
+ {\cal A}_{\chi_1^+ \chi_2^-}^{Z\g}, \nonumber\\
&&{\cal A}_{\chi_1^+\chi_2^-}^{Z\g}=
\frac{2 m_W}{\sqrt{
m_{\chi_1^\pm} m_{\chi_2^\pm}} } \bigg(
(g_{h\chi^+_1 \chi^-_2}^L g_{Z \chi^+_1 \chi^-_2}^L + g_{h\chi^+_1 \chi^-_2}^R g_{Z \chi^+_1 \chi^-_2}^R) A^{(1)}_{1/2},\nonumber\\
&& \quad \quad \quad \quad  \quad \quad \quad \quad  \quad + (g_{h\chi^+_1 \chi^-_2}^L g_{Z \chi^+_1 \chi^-_2}^R + g_{h\chi^+_1 \chi^-_2}^R g_{Z \chi^+_1 \chi^-_2}^L) A^{(2)}_{1/2}\bigg),
\end{eqnarray}
where $\lambda_i=4m_i^2/m_Z^2$ and the coupling coefficients of $h$ and $Z$ are given by
\begin{eqnarray}
&&\hat{v}_t=2T_3^{{t}}-4 Q_t s_w^2, \quad v_{H^\pm}=(c_w^2-s_w^2)/c_w, \\
&&v_{\tilde{f}_1}=(T_f^3 \cos^2\theta_{\tilde{f}}-Q_f s_w^2)/c_w, \quad\quad\quad\quad
v_{\tilde{f}_2}=(T_f^3 \sin^2\theta_{\tilde{f}}-Q_f s_w^2)/c_w, \\
&&v_{\tilde{f}_{12}}=(-T_f^3 \sin\theta_{\tilde{f}}\cos\theta_{\tilde{f}})/c_w, \\
&&g_{h\tilde f_1 \tilde f_2} = \frac{-1}{2(\sqrt{2}G_F)^{1/2}}\bigg(
\frac{1}{2}(\mathrm{Y}_{hRR}-\mathrm{Y}_{hLL})\sin2\theta_{\tilde f}+\mathrm{Y}_{hLR}\cos2\theta_{\tilde f}\bigg), \\
&&g_{Z \chi^+_1 \chi^-_1}^L=(V_{11}^2+1-2s^2_w)/(2\, c_w), \quad \quad\quad g_{Z \chi^+_1 \chi^-_1}^R=(U_{11}^2+1-2s^2_w)/(2\, c_w), \\
&&g_{Z \chi^+_2 \chi^-_2}^L=(V_{21}^2+1-2s^2_w)/(2\, c_w), \quad \quad\quad g_{Z \chi^+_2 \chi^-_2}^R=(U_{21}^2+1-2s^2_w)/(2\, c_w), \\
&&g_{Z \chi^+_1 \chi^-_2}^L=(V_{11}V_{21})/(2\, c_w),\quad \quad\quad\quad\quad\quad   g_{Z \chi^+_1 \chi^-_2}^R=(U_{11}U_{21})/(2\, c_w).
\end{eqnarray}

In above formula, we have defined some new functions as
\begin{eqnarray}
&&A^{(1)}_0=4 m_{\tilde{f}_1} m_{\tilde{f}_2} (C^{(1)}_{23}+C^{(1)}_{12})\nonumber\\
&&A^{(2)}_0=4 m_{\tilde{f}_1} m_{\tilde{f}_2} (C^{(2)}_{23}+C^{(2)}_{12})\nonumber\\
&&A^{(1)}_{1/2}=2 m_{\chi_1^\pm} \sqrt{ m_{\chi_1^\pm} m_{\chi_2^\pm}} \bigg( (2C^{(3)}_{23}+3C^{(3)}_{12}+C^{(3)}_0)+(2C^{(4)}_{23}+C^{(4)}_{12}) \bigg)\nonumber\\
&&A^{(2)}_{1/2}=2 m_{\chi_2^\pm} \sqrt{ m_{\chi_1^\pm} m_{\chi_2^\pm}} \bigg( (2C^{(4)}_{23}+3C^{(4)}_{12}+C^{(4)}_0)+(2C^{(3)}_{23}+C^{(3)}_{12})\bigg)
\end{eqnarray}
with $C_{ij}$ denoting three points loop functions introduced in \cite{loop}, and $C^{(1)}= C(P_\gamma,P_Z,m_{\tilde{f}_1},m_{\tilde{f}_1},m_{\tilde{f}_2})$, $C^{(2)}=C^{(1)}|_{m_{\tilde{f}_1}\leftrightarrow m_{\tilde{f}_2}}$, $C^{(3)}=C^{(1)}|_{m_{\tilde{f}_1} \rightarrow m_{\chi_1^\pm}, m_{\tilde{f}_2} \rightarrow m_{\chi_2^\pm}}$, and $C^{(4)}=C^{(3)}|_{m_{\chi_1^\pm}\leftrightarrow  m_{\chi_2^\pm} }$.
For the special cases considered here, $C_{ij}$ are given by
\begin{eqnarray}
&&C_0(p_\gamma,p_Z,m_1,m_1,m_2)=-\int^1_0 dy \frac{1}{b} \, \ln|\frac{by+c}{c}| \\
&&C_{12}(p_\gamma,p_Z,m_1,m_1,m_2)=\int^1_0 dy \frac{1-y}{b} \, \ln|\frac{by+c}{c}| \\
&&C_{23}(p_\gamma,p_Z,m_1,m_1,m_2)=\int^1_0 dy \frac{y(1-y)}{b} \bigg(1-
\frac{b+c}{b y}\,  \ln|\frac{by+c}{c}| \bigg)
\end{eqnarray}
where
\begin{eqnarray}
b=- (m_h^2-m_Z^2) (1-y), \quad c= -m_Z^2 y(1-y) + m_1^2 y + m_2^2(1-y).
\end{eqnarray}
Note that for the special case $m_1 =m_2$, these functions can be further simplified to get their analytic expressions. The other functions
relevant to our calculation are defined by
\begin{eqnarray}
A_{0}(x,y)&=& I_1(x,y),\\
A_{1/2}(x,y)&=& I_1(x,y)-I_2(x,y),\\
A_1(x,y)&=& 4 (3-\tan^2\theta_w) I_2(x,y)+ \left[ (1+2x^{-1}) \tan^2\theta_w - (5+2x^{-1})\right] I_1(x,y),
\end{eqnarray}
with
\begin{eqnarray}
 I_1(x,y) &=& \frac{x y}{2(x-y)} + \frac{x^2 y^2}{2(x-y)^2}[ f(x^{-1})-f(y^{-1})] + \frac{x^2 y}{(x-y)^2}[g(x^{-1})-g(y^{-1})],\\
 I_2(x,y) &=& - \frac{x y}{2(x-y)} [ f(x^{-1})-f(y^{-1})],\\
 g(x) &=& \sqrt{x^{-1} -1} \arcsin \sqrt{x}.
\end{eqnarray}

\end{document}